\documentclass[a4paper,fleqn,usenatbib]{mnras}

\usepackage{newtxtext,newtxmath}
\usepackage[T1]{fontenc}
\usepackage{ae,aecompl}

\usepackage{graphicx}	% Including figure files
\usepackage{amsmath}	% Advanced maths commands
\usepackage{hyperref}
\title[Relative planet occurrence rates]{Relative occurrence rates of terrestrial planets orbiting FGK stars} % Inffered from Kepler DR25}
\author[Sheng Jin]{
Sheng Jin$^{1}$\thanks{E-mail: shengjin@pmo.ac.cn}
\\
$^{1}$CAS Key Laboratory of Planetary Sciences, Purple Mountain Observatory, Chinese Academy of Sciences, Nanjing 210023, China
}

\date{Accepted XXX. Received YYY; in original form ZZZ}

\pubyear{2015}

\begin{document}
\label{firstpage}
\pagerange{\pageref{firstpage}--\pageref{lastpage}}
\maketitle

\begin{abstract}
This paper aims to derive a map of relative planet occurrence rates that can provide constraints on the overall distribution of terrestrial planets around FGK stars.
Based on the planet candidates in the Kepler DR25 data release, I first generate a continuous density map of planet distribution using a Gaussian kernel model and correct the geometric factor that the discovery space of a transit event decreases along with the increase of planetary orbital distance.
Then I fit two exponential decay functions of detection efficiency along with the increase of planetary orbital distance and the decrease of planetary radius. % using MCMC and EM algorithm.
Finally, the density map of planet distribution is compensated for the fitted exponential decay functions of detection efficiency to obtain a relative occurrence rate distribution of terrestrial planets.
The result shows two regions with planet abundance: one corresponds to planets with radii between 0.5 and 1.5 $R_{\oplus}$ within 0.2 AU, the other corresponds to planets with radii between 1.5 and 3 $R_{\oplus}$ beyond 0.5 AU.
It also confirms the features that may be caused by atmospheric evaporation: there is a vacancy of planets of sizes between 2.0 and 4.0 $R_{\oplus}$ inside of $\sim$ 0.5 AU, and a valley with relatively low occurrence rates between 0.2 and 0.5 AU for planets with radii between 1.5 and 3.0 $R_{\oplus}$.
\end{abstract}


%%%%%%%%%%%%%%%%%%%%%%%%%%%%%%%%%%%%%%%%%%%%%%%%%%

%%%%%%%%%%%%%%%%% BODY OF PAPER %%%%%%%%%%%%%%%%%%

\section{Introduction}

The catalog of known exoplanets and planet candidates has grown to $\sim$ 6000 since the discovery of 51 Pegasi b \citep{Mayor1995}.
Benefit from the progress of different detection methods \citep{Cumming2008,Mayor2011,Borucki2011,Cassan2012}, it has been clear that the presence of planets is common around FGK and M dwarf stars \citep{Gould2010,Howard2012,Bonfils2013,Dong2013,Kopparapu2013,Dressing2015,Gaidos2016}.
The mass, radius distributions and orbital properties of exoplanets become key tools for testing theories of planet formation and evolution \citep{Ida2004,Alibert2005,Ida2008,Mordasini2009,Youdin2011,Mordasini2012,Bitsch2015}.
 
An important link between exoplanet observation and theoretical research is the planet occurrence rate distribution that tells the frequencies of different types of planets per star \citep[e.g.,][]{Howard2010,Catanzarite2011,Fressin2013,Petigura2013,Dong2013,Burke2015,Fulton2017}.
These studies provide a deep understanding of the characteristics of the outcomes of planet formation and evolution.
For example, research has shown that the first discovered hot Jupiters are only a small proportion of planets, and most of the exoplanets are terrestrial planets and Jupiter-mass planets at large distances \citep{Cassan2012,Bonfils2013}.

To characterize the occurrence rates of planets is the main goal of NASA's Kepler mission \citep{Borucki2010,Koch2010,Batalha2013}.
Based on a four-year searching for transiting planets by recording the brightness of $\sim$ 2,000,000 stars, Kepler detected more than 4000 planet candidates, 2279 of which were confirmed as planets \citep{Thompson2018}.
This is the best sample so far for studying the planet occurrence rate. 
A lot of research has provided constraints on the frequency of Kepler-like planets \citep{Zhu2018,vanSluijs2018,He2019,Hsu2019,Hardegree2019,Yang2020}, even planets in the habitable zone \citep{Traub2012,vanSluijs2018,Mulders2018,Bryson2021}.
The occurrence rates of terrestrial planets at close-in orbits also reveal features that may be due to atmospheric evaporation, i.e., there is a deficit of Kepler planet candidates of sizes between 1.5 and 2.0 $R_{\oplus}$ \citep{Fulton2017,Kunimoto2020}.

Whether a transit event occurs depends on the geometric alignment of a planetary orbit, and most planets will not produce transit lightcurves \citep{Borucki1984}.
Thus an accurate model for the completeness of the Kepler pipeline \citep{Howard2012,Fressin2013} or an independent detection pipeline \citep{Petigura2013,Foreman-Mackey2014,Dressing2015} is needed to derive the occurrence rates for planets of different types per star.
Moreover, the probability to successfully discover a transit event through data analysis, i.e., the detection efficiency, decreases along the positive orbital distance axis and the negative planetary radius axis due to the decrease in the signal-to-noise ratio. 
This means that smaller planets with longer orbital distances are harder to discover in a transit survey.
Also, other important factors can affect the detection efficiency, like the reliability of a transit event \citep{Bryson2020}, the planet multiplicity on completeness \citep{Zink2019,Zhu2020}, etc.
Recently, approximate Bayesian computation has been adopted to compute the occurrence rates in a two-dimensional grid of orbital distance and planetary radius \citep{Hsu2019,Kunimoto2020}.

This work aims to derive a relative occurrence rate distribution that is normalized in a way that the maximum value is 1, rather than an absolute occurrence rate distribution that gives the frequency of the planet per star.
Such a relative occurrence rate distribution does not need an accurate model of the completeness of the Kepler pipeline.
It can provide the dense and sparse areas of planetary distribution, which is useful for comparing with the outcomes of the planet formation and evolution theory. 
For example, to compare the relative abundance of close-in terrestrial planets and the habitable zone planets \citep[e.g.,][]{Mayor2011,Traub2012,vanSluijs2018,Zhu2018}, to verify the bimodal distribution of planetary radius as predicted by atmospheric evaporation model \citep{Owen2013,Lopez2013,Jin2014,Fulton2017,Mordasini2020}, etc.

The reasoning behind the relative occurrence rate distribution is that there are only two factors needed to compensate for in calculating it. 
One is the decrease of the detection space at a larger orbital distance that is related to the geometric alignment of a planetary orbit, the other is the decrease of the detection efficiency at a larger orbital distance and at a smaller planetary radius.
The first is a geometric factor that can be accurately calculated, and the second factor can be modeled by exponential decay functions along the positive orbital distance axis and the negative planetary radius axis.
One major assumption in this paper is that the terrestrial planet distribution is uniform in a lot of subspaces in the orbital distance versus planetary radius space.
Therefore, by dividing the orbital distance versus planetary radius space into many subspaces and fitting the corresponding exponential decay function in each subspace, the statistical distribution of the exponential decay constant can be obtained.
To summarize this entire process, I first correct the geometric effect using the orbital inclinations, orbital distances, and the false-positive probabilities of the Kepler KOI (Kepler Object of Interest) planet candidates, and obtain a relative distribution density map of terrestrial planets.
Then, I fit two independent exponential decay functions along the positive orbital distance axis and the negative planetary radius axis using the Markov Chain Monte Carlo (MCMC) and expectation-maximization (EM) algorithms.
Finally, I compensate the relative distribution density of terrestrial planets for the exponential decay functions to derive the relative occurrence rate distribution.

This paper is organized as follows. 
Section \ref{sec:methods} describes the model used to compensate for the geometric effect and the detection efficiency.
Section \ref{sec:results} shows the fitted exponential decay constant of detection efficiency and the derived relative occurrence rate distribution of terrestrial planets.
Section \ref{sec:discuss} discusses the reliability of the assumption that the terrestrial planet distribution is uniform in a lot of subspaces.
Section \ref{sec:concl} summarizes the main findings of the relative occurrence rate distribution.

\section{Methods}

\label{sec:methods}

\subsection{Data and Gaussian Kernel Density}

The probability of an exoplanet to transit its star and then be discovered in a detection pipeline is mainly determined by its orbital inclination, orbital eccentricity, orbital distance, and its radius.
The inclination and eccentricity of a planet's orbit determines whether a transit event can occur, and in this paper it is called the geometric effect.
For simplicity of analysis, this work assumes that all the planet candidates are in circular orbits.
The effect of orbital eccentricity on the results will be discussed in Section \ref{sect:geo}.
The planet's orbital distance and the planetary radius determine the probability that a transit signal can be detected and successfully discovered by data analysis, and in this work the decrease of signal-to-noise ratio along the positive orbital distance axis and the negative planetary radius axis is called the detection efficiency.

\begin{figure}
	\includegraphics[width=1.05\columnwidth]{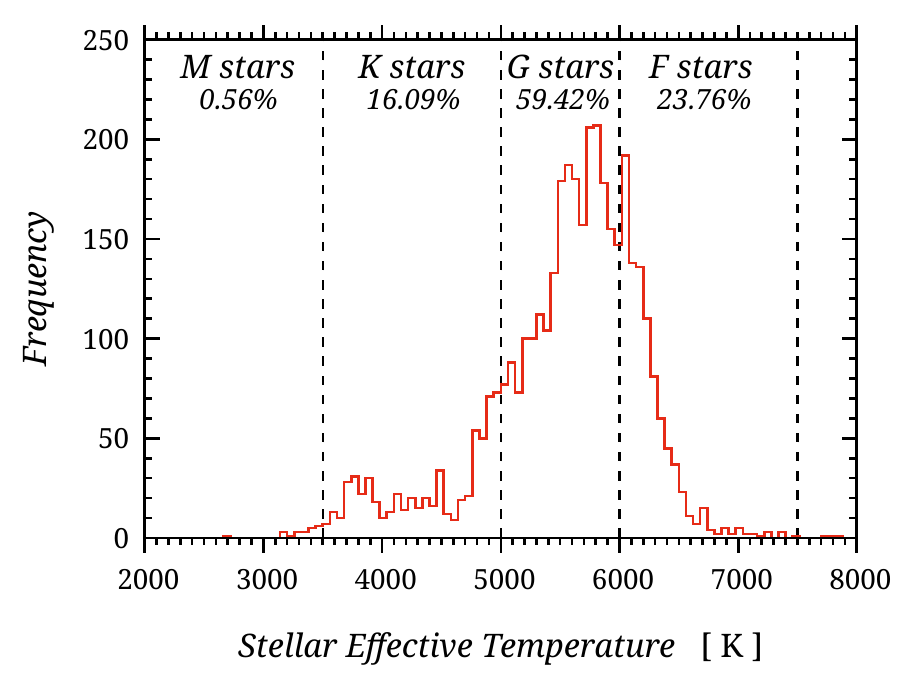}
    \caption{
	Histogram of the effective temperature of the host stars of all the 3928 samples used in this study. The labels in the picture indicate the corresponding spectral types of these stars.
    }
    \label{fig:1}
\end{figure}

Fortunately, the Kepler cumulative KOI table \footnote{\url{https://exoplanetarchive.ipac.caltech.edu/cgi-bin/TblView/nph-tblView?app=ExoTbls&config=cumulative}} provides all the information needed to quantify these effects, and it is in a dimensionless form.
To quantify and compensate for the geometric effect,
the orbital inclination, orbital distance and radius of a planet candidate need to be known (see Section \ref{sect:geo}).
These items are given by the columns of ``koi\_incl'' (Inclination [deg]), ``koi\_dor'' (Planet-Star Distance over Star Radius, hereafter $dor$) and ``koi\_ror'' (Planet-Star Radius Ratio, hereafter $ror$) in the Kepler cumulative KOI table.
To model the detection efficiency of a transit signal along the positive orbital distance axis and the negative planetary radius axis, the orbital distance and radius of a planet candidate need to be known (see Section \ref{sect:bias}), and these items are the save data that given by the ``koi\_dor'' and ``koi\_ror'' columns in the cumulative KOI table.

The Kepler false-positive probabilities table \footnote{\url{https://exoplanetarchive.ipac.caltech.edu/cgi-bin/TblView/nph-tblView?app=ExoTbls&config=koifpp}} provides the probability of each planet candidate to be any of the considered astrophysical false-positive signals, i.e., the ``fpp\_prob'' column in the table.
To rate the credibility of each planet candidate, the information given by the ``fpp\_prob'' column is also used to adjust the contribution weight of each planet candidate in the analysis.

There are a total of 9564 planet candidates in the cumulative KOI table, among which a total of 4617 have both $dor$ and $ror$ information.
4089 of these 4617 planet candidates have a given false-positive probability value.
161 false-positive signals with ``fpp\_prob'' equal to 1 were removed, 
leaving a total of 3928 samples.
Figure \ref{fig:1} shows the distribution of the effective temperature of the host stars of the 3928 planet candidates used in this study. 
The labels in the picture indicate the corresponding spectral types of the host stars.
It shows that M-type stars account for 0.56\%, K-type stars account for 16.09\%, G-type stars account for 59.42\%, and F-type stars account for 23.76\%.
In order to increase the sample population as much as possible, the 0.56\% candidates surrounding M-type stars are also added to the analysis.

\begin{figure}
	\includegraphics[width=1.05\columnwidth]{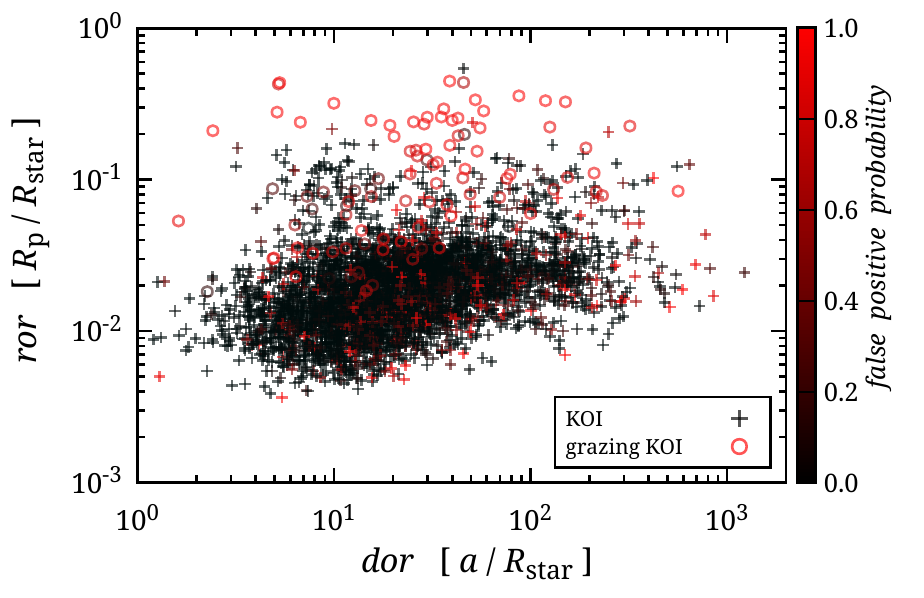}
    \caption{The $dor$-$ror$ distribution of the planet candidates in Kepler DR25 is shown,
	with circles showing the grazing candidates.
    The color of each point indicates the probability of a candidate to be any astrophysical false-positive scenarios.}
    \label{fig:2}
\end{figure}

Figure \ref{fig:2} shows all the planet candidates in the $dor$ versus $ror$ ($dor$-$ror$) space, with the color of each point giving the false-positive probability of each candidate. 
The circles in Figure \ref{fig:2} are the planet candidates with grazing incidence transits occurring, i.e., their transit event occurs at the stellar edge (see Section \ref{sect:geo}).
Most of the grazing candidates have a large false-positive probability, 
and they are excluded from the analysis because only their minimum planetary radii can be derived.

I use a Gaussian kernel density estimator \citep{Bishop2006} to convert the discrete distribution of planet candidates in the $dor$-$ror$ space into a continuous distribution that gives 
the dense and sparse regions of planetary distribution.  
The $dor$-axis and the $ror$-axis are set to linear coordinates, and the entire coordinate grid is represented by ${\mathbf x}$.
I place a two-dimensional Gaussian kernel density profile at the coordinate position of each planet candidate, and use the false-positive probability of the candidate to adjust the contribution weight of its Gaussian kernel density profile.
Then, a continuous distribution in the $dor$-$ror$ space is obtained by adding up the Gaussian density contributions of all the candidates at each grid point:
\begin{equation}
	\rho({\mathbf x}) = \sum_{n=1}^{N}\frac{1}{(2\pi h^2)^{^{\scriptscriptstyle 1/2}}} \exp \left\{ -\frac{ {\|{\mathbf x}-{\mathbf x}_{\scriptscriptstyle n}\|}^{\scriptscriptstyle 2} } {2h^{\scriptscriptstyle 2}} \right\} (1-fpp({\mathbf x}_{\scriptscriptstyle n}))
	\label{eq:gaus_kern}
\end{equation}
where the coordinate of a planet candidate is represented by ${\mathbf x}_{\scriptscriptstyle n}$,  $h$ represents the standard deviation of the two-dimensional Gaussian kernel, $fpp({\mathbf x}_{\scriptscriptstyle n})$ is the probability of a planet candidate to be any false-positive signals.

The only parameter in a Gaussian kernel density model is the kernel size $h$ that acts as a smoothing parameter.
It determines the dispersion of the Gaussian density profile, i.e., how the occurrence of a planet candidate is smoothed in adjacent space in $\mathbf x$.
There is a trade-off between being sensitive to noise with a small $h$ and being over-smoothed and washing out of the distribution characteristics with a large $h$.
Since the coordinate grid $\mathbf x$ is not dimensionless, I manually adjust $h$ to obtain a value that can smooth the appearance of a planet candidate properly in $\mathbf x$.
In addition to the fiducial kernel size $h$, this paper also presents two comparison cases with one fifth and five times the fiducial kernel size.
It shows that the final result of this model is not sensitive to the choice of $h$.
Because the distribution of all the planet candidates in $\mathbf x$ is fixed and thus the main features of the obtained distribution do not change with $h$.
Moreover, the detection efficiency needs to be fitted and then be compensated for is mainly due to the change of signal-to-noise ratio along the $dor$ and $ror$ axes, and that also does not change with $h$.

\subsection{Correction of Geometric Effect}
\label{sect:geo}

\begin{figure}
	\includegraphics[width=\columnwidth]{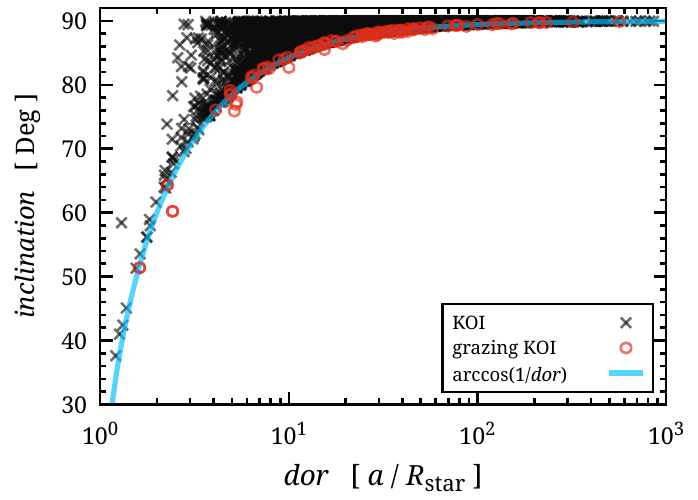}
    \caption{The $dor$-$i$ distribution of all the planet candidates.
	The blue line shows the minimum inclination where a transit event can occur in 
	the simplest case assuming a planet in a circular orbit.
	The red circles show the grazing planet candidates, 
	and they are excluded from the analysis since in these cases only 
	the minimum planetary radii can be derived.}
    \label{fig:3}
\end{figure}

\begin{figure*}
	\includegraphics[width=7.1in]{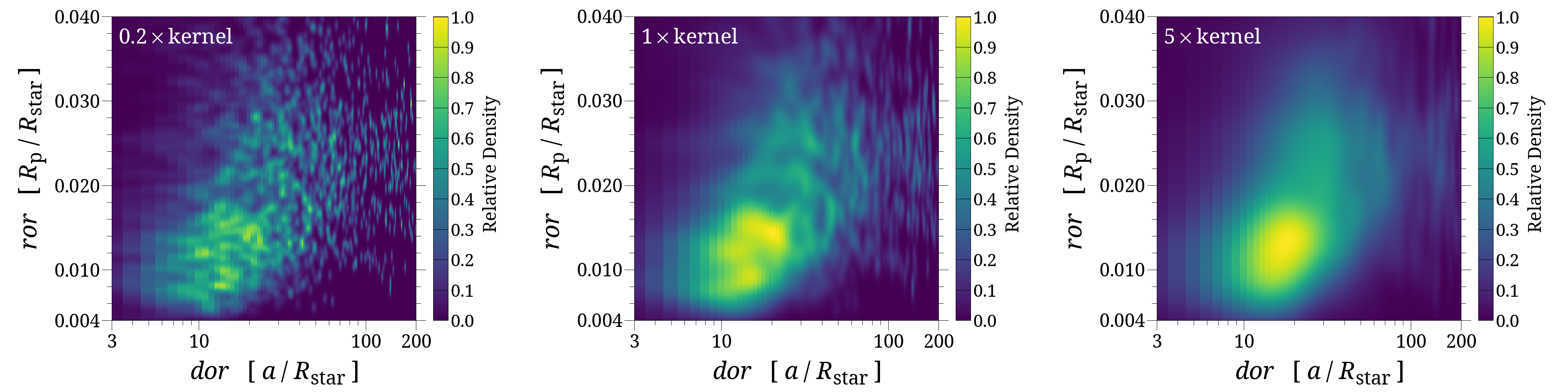}
	\caption{The relative distribution density of planet candidates after the correction of the decrease in the discovery space along the positive $dor$-axis for different $h$. The density profile in each panel is normalized by the maximum value in that case.}
    \label{fig:4}
\end{figure*}

The next step is to correct the decrease in the discovery space of transit events along the $dor$-axis due to the increase of the minimum inclination, $i_{\rm min}$, that determines whether a transit event can occur. 
In the simplest case assuming a planet in a circular orbit, $i_{\rm min}$ is given by:
\begin{equation}
	{\cos} \, i_{\rm min} = R_{\star} / a  = 1 / dor 
	\label{eq:incl_min}
\end{equation}
where $R_{\star}$ is the radius of a host star. There are also situations with grazing incidence transits occur:
\begin{equation}
	{\cos} \, i > \frac{R_{\star} - R_{\rm p}}{a} =  \frac{1-ror}{dor}
	\label{eq:incl_graz}
\end{equation}
where $R_{\rm p}$ is the radius of a planet candidate.
Since only the minimum planetary radii can be derived for the grazing planet candidates, these data are excluded from the analysis.

Figure \ref{fig:3} shows the orbital inclinations of all the 3928 planet candidates used in the analysis. It also shows the grazing planet candidates and the $i_{\rm min}$ given by Equation \ref{eq:incl_min}.
It shows clearly that the discovery space that lies between $i_{\rm min}$ and $90^{\circ}$ decreases rapidly along the positive $dor$-axis.
In Figure \ref{fig:3}, the orbital inclination distribution of all the planet candidates evenly fill the area between $i_{\rm min} < i < 90^{\circ}$. 
The uniformity of the distribution between $i_{\rm min}$ and $90^{\circ}$ implies that it is reasonable to scale with the angle interval along the positive $dor$-axis to compensate for the decrease of the discovery space due to the geometric effect, assuming exoplanets are evenly distributed in the 360-degree space of orbital inclination.

The $ror$ coordinate is also included here to calculate the change of $i_{\rm min}$ with planetary radius when grazing incidence transit occurs.
Again, in the simplest case of a circular orbit, the derived correction function for the decrease of discovery space is:
\begin{equation}
	corr(dor,ror) = \arccos(0) - \arccos(\frac{1-ror}{dor})
	\label{eq:f_corr}
\end{equation}
Correspondingly, the distribution density of planet candidates in the $dor$-$ror$ space that corrects the change in the discovery space along the positive $dor$-axis can be obtained by:
\begin{equation}
	\rho_{\scriptscriptstyle corr}({\mathbf x}) = \rho({\mathbf x}) / corr({\mathbf x})
	\label{eq:gaus_kern_corr}
\end{equation}

Note that using Equation \ref{eq:f_corr} derived in the case of a circular orbit is a major simplification in this work.
Although the multiple planet systems are on nearly circular (mean eccentricity $\bar{e}=0.04^{+0.03}_{-0.04}$) and coplanar orbits \citep{Xie2016}, especially for low-mass ones \citep{VanEylen2015}, Kepler singles are mostly on eccentric orbits of $\bar{e} \approx 0.3$ \citep{Xie2016}.
For eccentric orbits, the transit probability is a function of eccentricity, argument of pericentre $\omega$, and orbital semi-major axis.
Orbital eccentricity affects the probability of transit detection in two opposing ways: an increase in the probability for the 
planet to generate a transit near pericentre, 
accompanied by a reduction in the detectability due to a shorter transit duration \citep{Barnes2007,Burke2008}.
Integrating over the extent of the planet's shadow for all values of $\omega$ shows that planets on eccentric orbits are more likely to transit than planets on circular orbits \citep{Barnes2007,Kane2008,Burke2008}.
However, the enhancement of the transit probability is limited in most cases.
For planets with orbital periods $>$ 10 days, the decrease of the minimum orbital inclination $i_{min}$ for transiting planets is less than 10\% for eccentricity less than 0.7 \citep{Kane2008}.
For an idealized transit survey with a Rayleigh distribution of orbital eccentricities matching known planets, the calculation gives an enhancement of 4\% of the overall yield of transiting planet \citep{Burke2008}.
Thus, the simplification that assuming circular orbits has a limited effect on the final results.

Figure \ref{fig:4} shows the relative distribution density profiles of planet candidates after the correction of the decrease of the discovery space for different $h$.
The distribution density profiles shown in the figure is normalized by the maximum distribution density in each case.
The distribution density in the case of the fiducial kernel size  appeared as a smooth profile in the coordinate grid $\mathbf x$.
In the one-fifth kernel size case, the calculated distribution density profile presented in the form of many small clusters, because such a small $h$ cannot smooth out the sparsity of planet candidates in $\mathbf x$.
In the case with a five times kernel size, the sub-structures shown in the fiducial kernel size case were also smoothed out due to a large $h$.
However, in Section \ref{sec:results:size} it shows that the exponential decay functions of detection efficiency fitted in the three cases are similar.

\subsection{Correction of Detection Efficiency}
\label{sect:bias}

After obtaining the relative planet distribution density profile with compensation for the decrease of discovery space along the positive $dor$-axis, i.e., $\rho_{\scriptscriptstyle corr}({\mathbf x})$ in Equation \ref{eq:gaus_kern_corr}, I will fit two exponential decay functions of detection efficiency along the $dor$ and $ror$ axes.
The exponential decay function arises whenever a quantity decays at a rate proportional to its current value.
The weakening ability to discover planets along the positive $dor$-axis and the negative $ror$-axis exactly meets this characteristic. 

The exponential decay functions used in our model are:
\begin{equation}
	\widetilde{\rho}_{\scriptscriptstyle corr}({\mathbf x_{\scriptscriptstyle dor}}) = \widetilde{\exp}(-{\mathbf x}_{\scriptscriptstyle dor}/\tau_{\scriptscriptstyle dor})
	\label{eq:taudor}
\end{equation}
and 
\begin{equation}
	\widetilde{\rho}_{\scriptscriptstyle corr}({\mathbf x_{\scriptscriptstyle ror}}) = \widetilde{\exp} (-{\mathbf x}_{\scriptscriptstyle ror}/\tau_{\scriptscriptstyle ror})
	\label{eq:tauror}
\end{equation}
where $\widetilde{\rho}_{\scriptscriptstyle corr}({\mathbf x_{\scriptscriptstyle dor}})$ and $\widetilde{\rho}_{\scriptscriptstyle corr}({\mathbf x_{\scriptscriptstyle ror}})$ are the $dor$-axis and $ror$-axis components of $\rho_{\scriptscriptstyle corr}({\mathbf x})$ and they are normalized to sum to 1,
${\mathbf x}_{\scriptscriptstyle dor}$ and ${\mathbf x}_{\scriptscriptstyle ror}$ are the $dor$ and $ror$ coordinates in ${\mathbf x}$,
$\tau_{\scriptscriptstyle dor}$ and $\tau_{\scriptscriptstyle ror}$ are the decay constants that need to be fitted.
The exponential decay functions are also normalized to sum to 1, in order to be compared with the density profiles of $\widetilde{\rho}_{\scriptscriptstyle corr}({\mathbf x_{\scriptscriptstyle dor}})$ and $\widetilde{\rho}_{\scriptscriptstyle corr}({\mathbf x_{\scriptscriptstyle ror}})$.

An important assumption have been made by using Equation \ref{eq:taudor} and \ref{eq:tauror} to fit the distribution of planet candidates, that is the distribution of exoplanets is mostly uniform in the $dor$-$ror$ space.
While such an assumption is inaccurate.
For example, theoretical studies have predicted the planetary radius distribution should be bimodal due to the effect of atmospheric evaporation, a feature that is latter shown in the planet occurrence rate derived by Kepler data \citep{Owen2013,Jin2014,Fulton2017,Kunimoto2020}. 
Research also shows that the planet occurrence rate distribution is not uniform along the orbital distance axis \citep{Traub2012,Dong2013,vanSluijs2018,Mulders2018}.
The actual distribution of the observed exoplanets should be a combination of unknown features and an overall trend of exponential decay.
Here, the strategy is to derive the exponential decay functions statistically, assuming that the distribution of terrestrial planets is uniform in most subspaces of ${\mathbf x}$.
A detailed discussion of this assumption will be presented in Section \ref{sec:discuss}.
I divide the grid ${\mathbf x}$ into many subspaces, and respectively fit $\tau_{\scriptscriptstyle dor}$ and $\tau_{\scriptscriptstyle ror}$ in each subspace using the maximum a posterior (MAP) estimator obtained by MCMC sampling, and then obtain the statistical distributions of $\tau_{\scriptscriptstyle dor}$ and $\tau_{\scriptscriptstyle ror}$.

If the actual distribution of exoplanets is uniform in ${\mathbf x}$ space, then the fitted $\tau_{\scriptscriptstyle dor}$ and $\tau_{\scriptscriptstyle ror}$ in all the subspaces will form a Gaussian distribution.
However, the local features of the realistic distribution given by the Kepler planet candidates will lead to many outliers in the distributions of $\tau_{\scriptscriptstyle dor}$ and $\tau_{\scriptscriptstyle ror}$.
In this case, the obtained distributions of $\tau_{\scriptscriptstyle dor}$ and $\tau_{\scriptscriptstyle ror}$ in all the subspaces will meet the Student's t-distribution \citep{Bishop2006}:
\begin{equation}
	St(\tau\vert\mu,\lambda,\nu) = \frac{\Gamma(\nu/2+1/2)}{\Gamma(\nu/2)} \left(\frac{\lambda}{\pi\nu}\right)^{1/2} \left[1+\frac{\lambda(\tau-\mu)^2}{\nu}\right]^{-\nu/2-1/2}
	\label{eq:st}
\end{equation}
where $\tau$ is $\tau_{\scriptscriptstyle dor}$ or $\tau_{\scriptscriptstyle ror}$,
$\mu$ and $\lambda$ are the mean and the precision of the Student's t-distribution, 
$\nu$ is the degrees of freedom, and $\Gamma$ is the Gamma function.

The EM algorithm is used to find the maximum likelihood estimates for the parameters in the Student's t-distribution of the fitted $\tau_{\scriptscriptstyle dor}$ and $\tau_{\scriptscriptstyle ror}$ in all the subspaces \citep{Scheffler2008}.
Then, I compensate the $\rho_{\scriptscriptstyle corr}({\mathbf x})$ for the exponential decay functions described by the mean values of the Student's t-distribution of $\tau_{\scriptscriptstyle dor}$ and $\tau_{\scriptscriptstyle ror}$, after which the relative occurrence rates of Kepler planet candidates in ${\mathbf x}$ can be obtained.
Besides, I also derive the relative planet occurrence rates using the $\tau_{\scriptscriptstyle dor}$ and $\tau_{\scriptscriptstyle ror}$ at $\pm 1\sigma$ locations to show the extreme cases.

\section{Results}
\label{sec:results}

\begin{figure*}
	\includegraphics[width=6.8in]{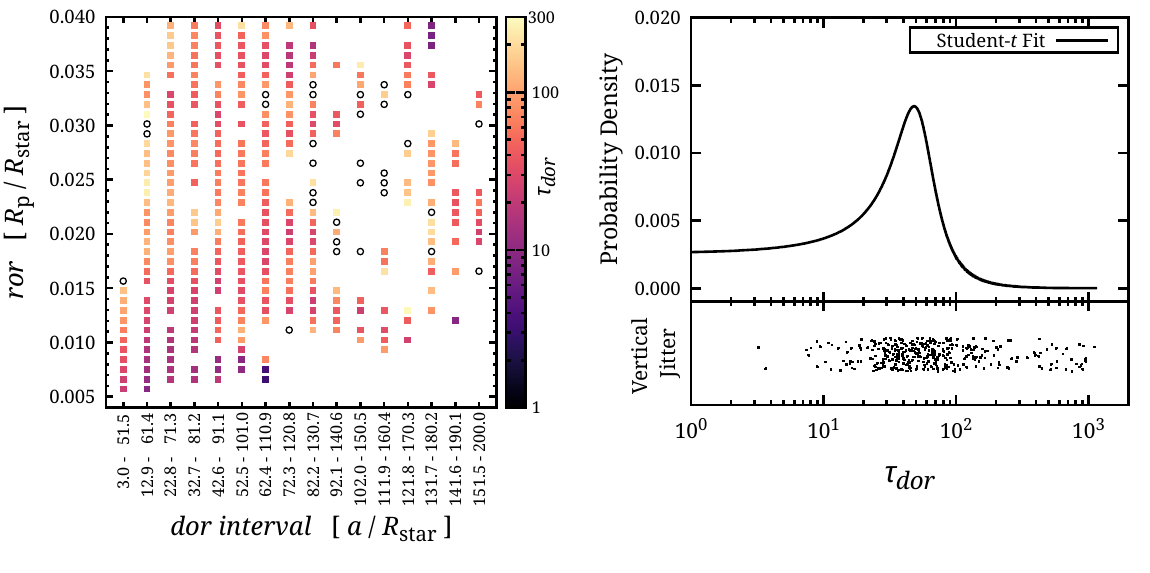}
	\caption{
	%Left plot: The MAP estimator of $\tau_{\scriptscriptstyle dor}$ obtained via MCMC sampling in the first and second categories, and they are used in the fitting of the Student's t-distribution using EM algorithm.
	Left plot: The fitted  $\tau_{\scriptscriptstyle dor}$  in the subspaces belong to the first and second categories, and these values are used in the fitting of the Student's t-distribution of $\tau_{\scriptscriptstyle dor}$ using EM algorithm.
	The color of each point shows the value of $\tau_{\scriptscriptstyle dor}$ in the subspaces belong to the first category.
	The black circles shows the subspaces belong to the second category that are characterized by a very large $\tau_{\scriptscriptstyle dor}$, corresponding to an extremely slow decay rate.
	Right plot: The top panel shows the fitted Student's t-distribution of $\tau_{\scriptscriptstyle dor}$ using the fitted values of the subspaces belong to the first and second categories, and the bottom panel shows all these  values. % of these $\tau_{\scriptscriptstyle dor}$.
	Note that vertical jitter has been added in the bottom graph to make it easier to see the individual samples.
	}
    \label{fig:5}
\end{figure*}

\begin{figure}
	\includegraphics[width=3.5in]{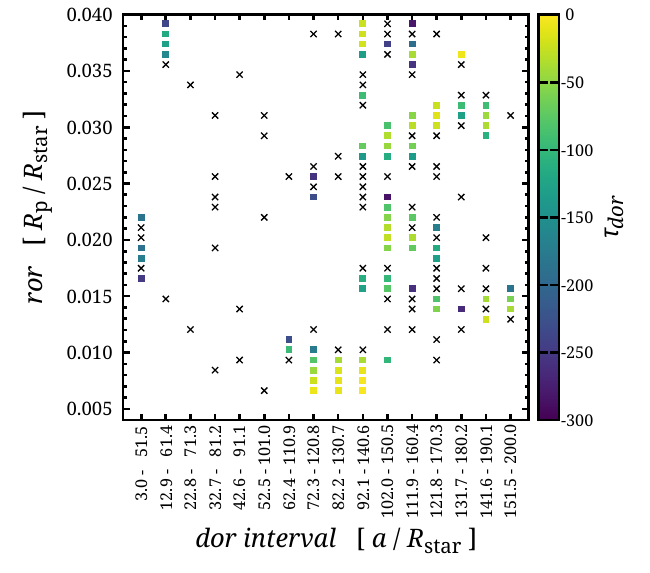}
	\caption{
	The color points show the fitted values of $\tau_{\scriptscriptstyle dor}$ in the subspaces belong to the third category, and these values correspond to distribution density profiles that increase along the positive $dor$-axis.
	The black crosses show the subspaces belong to the fourth category that are characterized by density profiles of multiple peaks that cannot be adequately fitted by an exponential function. 
	}
    \label{fig:6}
\end{figure}

\subsection{Distributions of $\tau_{\scriptscriptstyle dor}$}

In the fitting of $\tau_{\scriptscriptstyle dor}$ and $\tau_{\scriptscriptstyle ror}$, the ${\mathbf x}$ is selected as the space with the $dor$-axis from 3 to 200 and the $ror$-axis from 0.004 to 0.04.
The $dor$-axis and $ror$-axis are linearly divided into 200 grid points.
In the case of $\tau_{\scriptscriptstyle dor}$, 16 partially overlapping segments are extracted along the positive $dor$-axis, with 50 grid points in each segment and a stride size of 10 grids in the $dor$-axis.
An average relative distribution density, $\overline{\rho}_{\scriptscriptstyle corr}$, is calculated for every ten adjacent grid points on the $ror$-axis, and the stride size in the $ror$-axis is set as 5 grids.
By this, I divide the space ${\mathbf x}$ into 624 subspaces, $16\times39$, and each subspace gives a density profile of 50 grid points along the positive $dor$-axis.
Note that those subspaces where the density profiles that are clearly characterized by insufficient detection data are skipped.
In practice, I only fit the $\tau_{\scriptscriptstyle dor}$ of the subspaces lie between the leftmost and rightmost grid points that are greater than a critical relative distribution density ${\rho}_{\scriptscriptstyle corr\_crit}$, where the ${\rho}_{\scriptscriptstyle corr\_crit}$ is set as the $1/35$ of the maximum relative distribution density in the space ${\mathbf x}$.
Combining Figure \ref{fig:5} and Figure \ref{fig:6}, these skipped subspaces are located in the upper left corner, upper right corner, and lower right corner of the $dor$-$ror$ space.

In each subspace, the $\tau_{\scriptscriptstyle dor}$ is fitted using the MAP estimator obtained via MCMC sampling.
There are four categories of the fitted $\tau_{\scriptscriptstyle dor}$.
The first category consists of subspaces characterized by a positive fitted $\tau_{\scriptscriptstyle dor}$ that is less than 300.
The second category consists of subspaces characterized by a positive fitted $\tau_{\scriptscriptstyle dor}$ that is greater than 300.
The density profiles of the subspaces in these two categories can be fitted with an exponential decay function.
The difference is that the subspaces in the second category have an extremely slow decay rate.
In reality, the detection efficiency should not decrease so slowly along the positive $dor$-axis, thus the density profiles of the subspaces in the second category may be related to local features or the outliers caused by insufficient detection. 
The third category consists of subspaces characterized by a negative value of $\tau_{\scriptscriptstyle dor}$, indicating that the actual density profiles in these subspaces increase along the positive $dor$-axis.
The fourth category consists of subspaces characterized by density profiles of multiple peaks that cannot be properly fitted by an exponential function, i.e., extremely uneven density distributions along the $dor$-axis.  
These two categories consist of subspaces with highly skewed density distributions, thus the fitted values of $\tau_{\scriptscriptstyle dor}$ in these subspaces are not used for deriving the degree of exponential decay of detection efficiency.
The left panel of the Figure \ref{fig:5} shows the distribution of the subspaces belong to the first and second categories, and Figure \ref{fig:6} shows the distribution of the subspaces belong to the third and fourth categories.
Most grid points with $dor$ < $\sim$ 100 correspond to the subspaces in the first category, which implies that the distribution profile of planets can be represented by an exponential decay function along the positive orbital distance axis at close-in orbits.
At grids with $dor$ > $\sim$ 100, the four categories are distributed chaotically, and they contain a similar number of subspaces.
This implies that the rich and varied local features at larger orbital distances may be caused by noise.

Based on the values of the first and second categories,  the maximum likelihood estimates for the parameters in the Student's t-distribution of $\tau_{\scriptscriptstyle dor}$ are fitted using the EM algorithm. 
The fitted $\mu$ in the Student's t-distribution is 48.21,
$\lambda$ and $\nu$ are 0.0018 and 0.97,
and the standard deviation $\sigma$ is 23.52.
This distribution is shown in the right plot in Figure \ref{fig:5}.
Since this fitting process of the exponential decay constant $\tau_{\scriptscriptstyle dor}$ relies on the assumption that exoplanets are uniformly distributed along the $dor$-axis in most subspaces, an additional discussion on its reliability will be presented in Section \ref{sec:discuss}.

\subsection{Distributions of $\tau_{\scriptscriptstyle ror}$}

\begin{figure*}
	\includegraphics[width=6.8in]{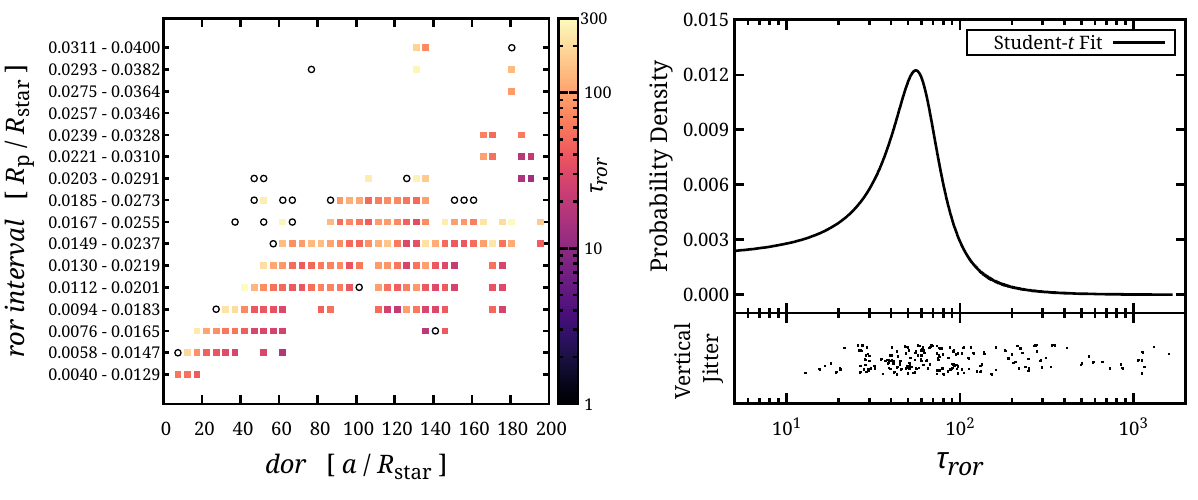}
	\caption{
	%Left plot: The MAP estimator of $\tau_{\scriptscriptstyle ror}$ obtained via MCMC sampling in the first and second categories, and they are used in the fitting of the Student's t-distribution using EM algorithm.
	Left plot: The fitted  $\tau_{\scriptscriptstyle ror}$ in the subspaces belong to the first and second categories, and these values are used in the fitting of the Student's t-distribution of $\tau_{\scriptscriptstyle ror}$ using EM algorithm.
	The color of each point shows the value of $\tau_{\scriptscriptstyle ror}$ in the subspaces belong to the first category.
	The black circles shows the subspaces belong to the second category that are characterized by a very large $\tau_{\scriptscriptstyle ror}$, corresponding to a very slow decay rate.
	Right plot: The top panel shows the fitted Student's t-distribution of $\tau_{\scriptscriptstyle ror}$ using the fitted values of the subspaces belong to the first and second categories. 
	The bottom panel shows all the values of these  $\tau_{\scriptscriptstyle ror}$, and vertical jitter has been added in the graph to make it easier to see the individual points.
	}
    \label{fig:7}
\end{figure*}

\begin{figure}
	\includegraphics[width=3.5in]{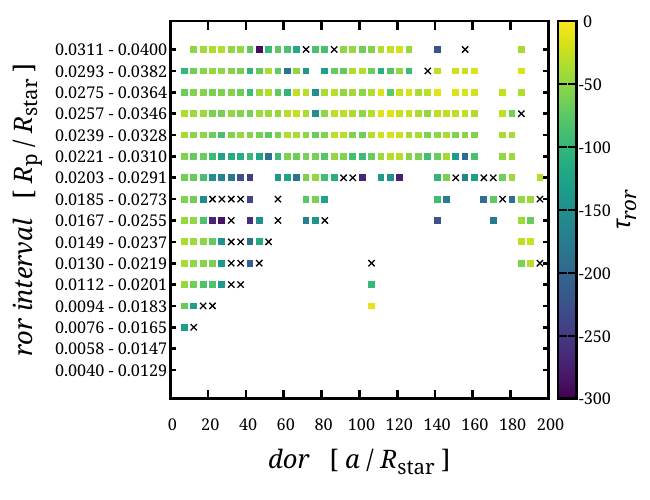}
	\caption{
	The color points show the fitted values of $\tau_{\scriptscriptstyle ror}$ in the subspaces belong to the third category, and these values correspond to distribution profiles that increase along the negative $ror$-axis.
	The black crosses show the subspaces belong to the fourth category that are characterized by multiple peaks that cannot be adequately fitted by an exponential function. 
	}
    \label{fig:8}
\end{figure}

In the case of $\tau_{\scriptscriptstyle ror}$, 16 partially overlapping segments are extracted along the negative $ror$-axis, with 50 grid points in each segment and a stride size of 10 grids in the $ror$-axis.
An average relative distribution density for every ten adjacent grids is calculated on the $dor$-axis, and the stride size in the $dor$-axis is 5 grids.
This also leads to 624 relative density profiles and each profile consists of 50 grid points along the negative $ror$-axis.
Similarly, the subspaces with density profiles that are characterized by insufficient detection data are skipped.
I only fit the $\tau_{\scriptscriptstyle ror}$ of the subspaces lie between the top and bottom grid points that are greater than the critical relative distribution density ${\rho}_{\scriptscriptstyle corr\_crit}$.
Combining Figure \ref{fig:7} and Figure \ref{fig:8}, these skipped subspaces are located in the upper right corner and a large area in the lower right corner of the $dor$-$ror$ space.

The $\tau_{\scriptscriptstyle ror}$ is fitted using the MAP estimator obtained via MCMC sampling in each subspace. 
Just like in the case of $\tau_{\scriptscriptstyle dor}$, there are also four categories of the fitted $\tau_{\scriptscriptstyle ror}$, and the four categories are defined identically to those of $\tau_{\scriptscriptstyle dor}$ except for the variable change.
The left plot in Figure \ref{fig:7} gives the distributions of the fitted $\tau_{\scriptscriptstyle ror}$ of the subspaces belong to the first and second categories.
Figure \ref{fig:8} shows the distributions of the subspaces belong to the  third and fourth  categories.
Figure \ref{fig:7} shows that most of the subspaces belong to the first category are located in the lower part of the $dor$-$ror$ space, corresponding to planets of sizes between $\sim$ 1-2 $R_{\oplus}$.
However, the largest category is the subspaces belong to the third one with a negative fitted $\tau_{\scriptscriptstyle ror}$, as shown in Figure \ref{fig:8}, and nearly all the subspaces in the upper part of the plot and the inner area belong to the third category.
This means that for planets of sizes between $\sim$ 2-4 $R_{\oplus}$, their distribution profile increases along the negative $ror$-axis, although the detection efficiency should decrease in this direction.
Such a distribution feature of $\tau_{\scriptscriptstyle ror}$ informs that there are significantly more terrestrial planets of sizes between 1-2 $R_{\oplus}$.
The number of the subspaces belong to the second and fourth categories are relatively small, and most of them are located at the transition area between the first and third categories.
The transition boundary between the first and third categories increases along the $dor$-axis in the $dor$-$ror$ space. 
This feature informs that the sizes of the majority of terrestrial planets increases as the positive orbital semi-major axis increases.

Based on the values of the first and second categories, the maximum likelihood estimates for the parameters in the Student's t-distribution of $\tau_{\scriptscriptstyle ror}$ are fitted using the EM algorithm.
The fitted $\mu$, $\lambda$, $\nu$, and the standard deviation $\sigma$ in the Student's t-distribution are 55.65, 0.0016, 0.82, and 24.96.
The right plot in Figure \ref{fig:7} shows this distribution.
A detailed discussion of this fitting process will be presented in Section \ref{sec:discuss}.

\subsection{Effect of Gaussian Kernel Size}
\label{sec:results:size}

\begin{table}
	\centering
	\caption{The parameters of the fitted Student's t-distributions of $\tau_{\scriptscriptstyle dor}$ and $\tau_{\scriptscriptstyle ror}$ for runs with different kernel sizes.}
	\label{tab:1}
	\begin{tabular}{ccccc} % four columns, alignment for each
		\hline
		kernel size & $\mu$ & $\lambda$ & $\nu$ & $\sigma$ \\
		(fiducial size) &  &  &  &  \\
		\hline
		\multicolumn{5}{|c|}{$\tau_{\scriptscriptstyle dor}$} \\
		\hline
		0.2 & 41.84 & 0.0016 & 0.94 & 24.93 \\
		1   & 48.21 &  0.0018 & 0.97 & 23.52 \\
		5 & 51.86 & 0.0029 & 0.83 & 18.66 \\
		\hline
		\multicolumn{5}{|c|}{$\tau_{\scriptscriptstyle ror}$} \\
		\hline
		0.2 & 48.56 & 0.0013 & 0.71 & 27.42 \\
		1   & 55.65 & 0.0016 & 0.82 & 24.96 \\
		5 & 57.51 & 0.0022 & 0.82  & 21.25 \\
		\hline
	\end{tabular}
\end{table}

\begin{figure*}
    \centering
    \begin{minipage}{0.5\textwidth}
        \centering
        \includegraphics[width=0.95\textwidth]{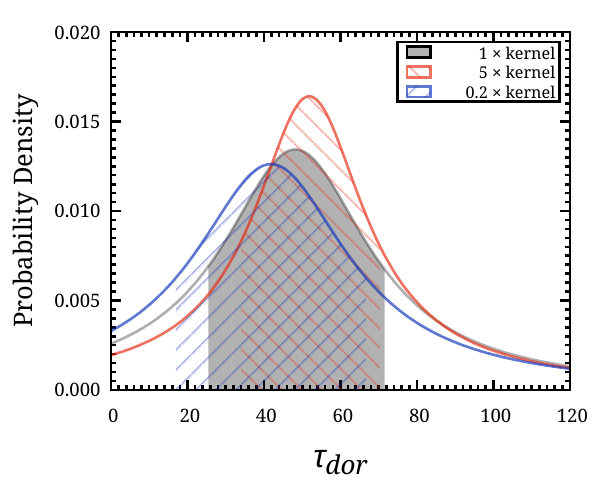}
        %\caption{first figure}
    \end{minipage}\hfill
    \begin{minipage}{0.5\textwidth}
        \centering
        \includegraphics[width=0.95\textwidth]{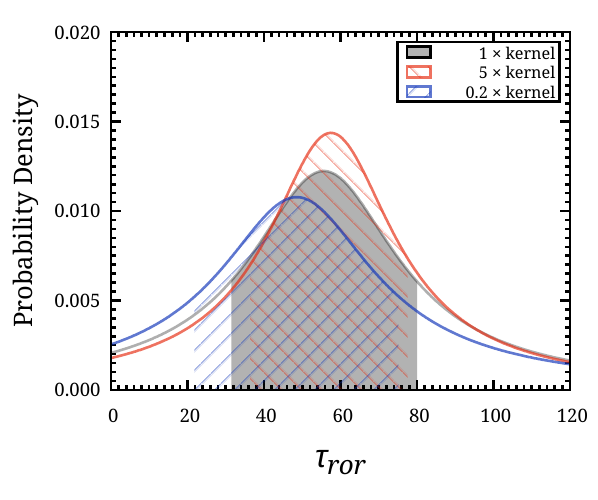} 
        %\caption{second figure}
    \end{minipage}
    \caption{Left plot: The Student's t-distributions of $\tau_{\scriptscriptstyle dor}$ 
	obtained using a kernel size $h$ with 0.2, 1, and 5 times the fiducial kernel size. 
    The shaded part of each curve gives the 1-$\sigma$ region of the fitted $\tau_{\scriptscriptstyle dor}$.
	Right plot: The Student's t-distributions of $\tau_{\scriptscriptstyle ror}$ 
	obtained in the three cases with different kernel size $h$,
	and the shaded part of each curve shows the 1-$\sigma$ region.}
    \label{fig:9}
\end{figure*}

Here I study the effect of the kernel size $h$ on the fitted $\tau_{\scriptscriptstyle dor}$ and $\tau_{\scriptscriptstyle ror}$.
As shown in Figure \ref{fig:4}, the relative distribution density after the correction of the geometric effect varies greatly in smoothness for different $h$, which are 0.2, 1, and 5 of the fiducial kernel size.
However, the overall trend of the distribution exhibited in the three cases are similar, because they are determined by the distribution characteristics of all the planet candidates in $\mathbf x$, which is fixed.
As a consequence, the Student's t-distributions of $\tau_{\scriptscriptstyle dor}$ and $\tau_{\scriptscriptstyle ror}$ obtained by the maximum likelihood estimates for the three cases do not change much.
Table \ref{tab:1} gives the fitted parameters obtained using different $h$.
For $\tau_{\scriptscriptstyle dor}$, the $\mu$ fitted under 0.2, 1, and 5 times the fiducial kernel size are 41.84, 48.21, and 51.86, and the $\sigma$ are 24.93, 23.52, and 18.66.
For $\tau_{\scriptscriptstyle ror}$, the $\mu$ under 0.2, 1, and 5 times the fiducial kernel size are 48.56, 55.65, and 57.51, and the $\sigma$ are 27.42, 24.96, and 21.25.

Figure \ref{fig:9} plots the Student's t-distributions of the fitted $\tau_{\scriptscriptstyle dor}$ and $\tau_{\scriptscriptstyle ror}$ for the three cases.
The probability density curves of both $\tau_{\scriptscriptstyle dor}$ and $\tau_{\scriptscriptstyle ror}$ obtained with these three kernel sizes are similar, and the 1-$\sigma$ regions of the three distributions are generally overlapping.
This confirms that the kernel size only acts as a smoothing factor, and it has little effect on the fitted exponential decay functions of detection efficiency. % obtained by the maximum likelihood estimates.
This also indicates that the relative occurrence rates of planet candidates derived using the fiducial kernel size in the next section are robust.

\subsection{Relative Planet Occurrence Rates}

Using the mean values of the Student's t-distribution fits of $\tau_{\scriptscriptstyle dor}$ and $\tau_{\scriptscriptstyle ror}$, the $\rho_{\scriptscriptstyle corr}({\mathbf x})$ is compensated for the exponential decay functions of detection efficiency described by Equation \ref{eq:taudor} and \ref{eq:tauror}.
In this way the relative occurrence rate distribution of planet candidates in ${\mathbf x}$ are derived.
Since the range of the $ror$-axis in ${\mathbf x}$ is from 0.004 to 0.04, the result corresponds to terrestrial planets. 

\begin{figure*}
	\includegraphics[width=7.0in]{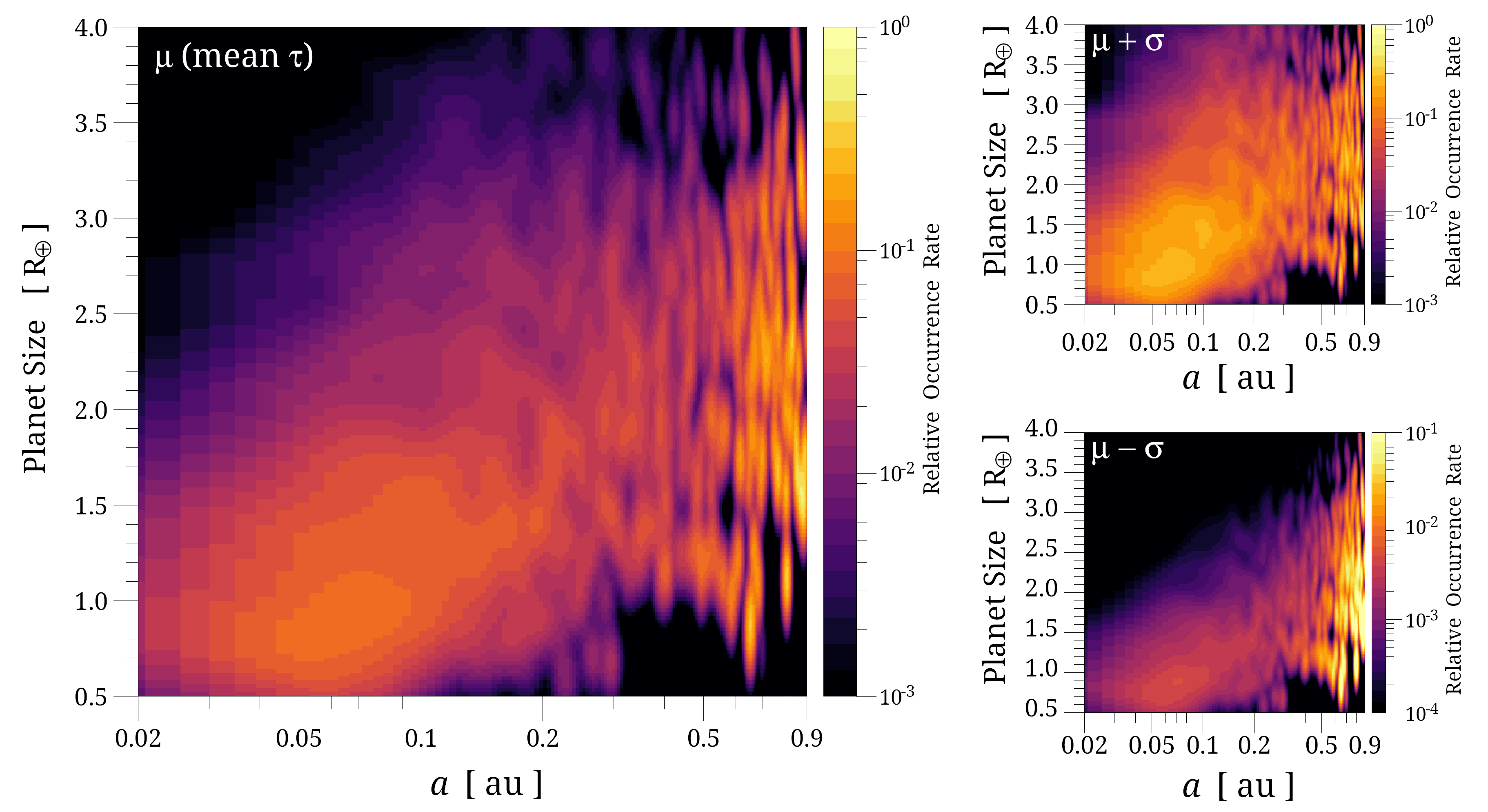}
	\caption{Left plot: The relative occurrence rates of terrestrial planets after the corrections of the geometric effect and the exponential decay of detection efficiency.
	Here the exponential decay functions are modeled using the mean values of $\tau_{\scriptscriptstyle dor}$ and $\tau_{\scriptscriptstyle ror}$ fitted using the fiducial kernel size.
	Right plots: The relative planet occurrence rates obtained using the values at one standard deviations above and below the mean ($\mu$+$\sigma$ for the top plot and $\mu$-$\sigma$ for the bottom) of $\tau_{\scriptscriptstyle dor}$ and $\tau_{\scriptscriptstyle ror}$ fitted using the fiducial kernel size.
	Note that in the bottom right plot, the highest value of the colorbar is set to 0.1 to highlight the areas with low relative planet occurrence rates.
	}
    \label{fig:10}
\end{figure*}

The left plot of Figure \ref{fig:10} gives the relative occurrence rate distribution derived using the mean values of the Student's t-distribution fits of $\tau_{\scriptscriptstyle dor}$ and $\tau_{\scriptscriptstyle ror}$ in the fiducial kernel size case.
There are two regions with very low planet occurrence rates as shown in this plot, one in the upper left corner and the other in the lower right corner.
The upper left corner is the area with the highest detection efficiency, so the reason for the vacancy of planet may be due to the effect of atmospheric escape \citep{Owen2013,Lopez2013,Jin2014,Chen2016}.
It shows clearly that as the orbital semi-major axis increases, the lower boundary of this vacancy region gradually increases, which exactly corresponds to the weakening of the stripping of planetary envelope due to atmospheric evaporation \citep{Owen2017,Lopez2018,Jin2018,Mordasini2020}.
The vacancy in the lower right corner is because no planet candidate has been detected there, so the relative occurrence rates cannot be increased even if the compensation factor there is high.
Thus, the shape of this vacancy region does not change with the variations of $\tau_{\scriptscriptstyle dor}$ and $\tau_{\scriptscriptstyle ror}$, as can be seen from the two comparison plots on the right of Figure \ref{fig:10}.

This plot also shows two regions with planet abundance.
One region corresponds to planets with radii between 0.5 and 1.5 $R_{\oplus}$  within 0.2 AU, indicating that there are a large number of close-in terrestrial planets \citep{Fressin2013}. 
Considering the effect of atmospheric escape, this also implies that most bare planetary cores are with radii of $0.5$-$1.5$ $R_{\oplus}$.
The other high occurrence rate region corresponds to planets with radii between 1.5 and 3 $R_{\oplus}$ beyond 0.5 AU.
This is the region with the highest occurrence rates of planets, with $\sim$ 10 times higher than the occurrence rates in the high occurrence rate region within 0.2 AU, suggesting the abundance of terrestrial planets in the habitable zone around FGK stars \citep{Traub2012,vanSluijs2018}.

One important region in this plot corresponds to planets with radii between 1.5 and 3.0 $R_{\oplus}$ and orbital semi-major axes between 0.2 and 0.5 AU. 
This is an area with considerable planet occurrence rates, but much smaller than the two regions with planet abundance.
It corresponds to the valley with low planet occurrence rates found by theoretical studies of atmospheric escape \citep{Owen2013,Lopez2013,Jin2014,Owen2017,Jin2018,Lopez2018}, a feature that was later confirmed by the planet occurrence rate distribution derived from Kepler data \citep{Fulton2017,Kunimoto2020}.
Compared with the area beyond 0.5 AU, the detection efficiency in this valley is much higher, but the planet occurrence rates here are only about one-tenth of the occurrence rates beyond 0.5 AU.
The distribution characteristics of this area are similar to those seen in Figure 2 of \citet{Hsu2019}.
Note that this is not a region with planet vacancy as shown in theoretical studies \citep{Owen2013,Jin2014}, as the relative planet occurrence rates in this region are still noticeable.
It is possible that an overall exponential decay function simplifies the actual decrease of the detection efficiency along the positive $dor$-axis and the negative $ror$-axis, and results in deviations in the local characteristics of the final relative occurrence rate distribution.
For example, if in reality the decay of detection efficiency is slower at close-in orbits and faster at large orbital distances, or slower at large planetary radii and faster at small planetary radii,
then the relative depletion of planets in the valley between 0.2 and 0.5 AU  will be more obvious after compensating for such decay functions.  

The one standard deviations above and below the mean values of the fitted Student's t-distribution of $\tau_{\scriptscriptstyle dor}$ and $\tau_{\scriptscriptstyle ror}$ are used to investigate the relative planet occurrence rates resulting from extremely fast or slow exponential decay functions of detection efficiency.
The top right plot in Figure \ref{fig:10} shows the occurrence rates derived by $\tau_{\scriptscriptstyle dor}$ and $\tau_{\scriptscriptstyle ror}$ at the one standard deviation above the mean, corresponding to the case of slower exponential decay functions of detection efficiency.
The regions with high and low relative planet occurrence rates are similar to the high and low occurrence rate regions obtained from the mean values of $\tau_{\scriptscriptstyle dor}$ and $\tau_{\scriptscriptstyle ror}$.
The main difference here is that the two high occurrence rate regions, one within 0.2 AU and the other beyond 0.5 AU, are in the same magnitude in this case.
The bottom right plot in Figure \ref{fig:10} shows the results derived by $\tau_{\scriptscriptstyle dor}$ and $\tau_{\scriptscriptstyle ror}$ at the one standard deviation below the mean, corresponding to faster exponential decay functions of detection efficiency.
The highest value of the colorbar is set to 0.1 in this plot to highlight the areas with low relative planet occurrence rates, because in this run the occurrence rates are too high in the outer high occurrence rate region beyond 0.5 AU.
In this case,  most of the terrestrial planets are with radii between 1.5 and 3 $R_{\oplus}$ and semi-major axes $>$ 0.5 AU.
The occurrence rates of planets with radii between 0.5 and 1.5 $R_{\oplus}$ inside of 0.2 AU are significantly lower, so that they are in the same magnitude as the occurrence rates in the region that corresponds to the valley of atmospheric evaporation.

\section{Discussion: Assumption of uniform distribution}
\label{sec:discuss}

\begin{figure}
	\includegraphics[width=1.05\columnwidth]{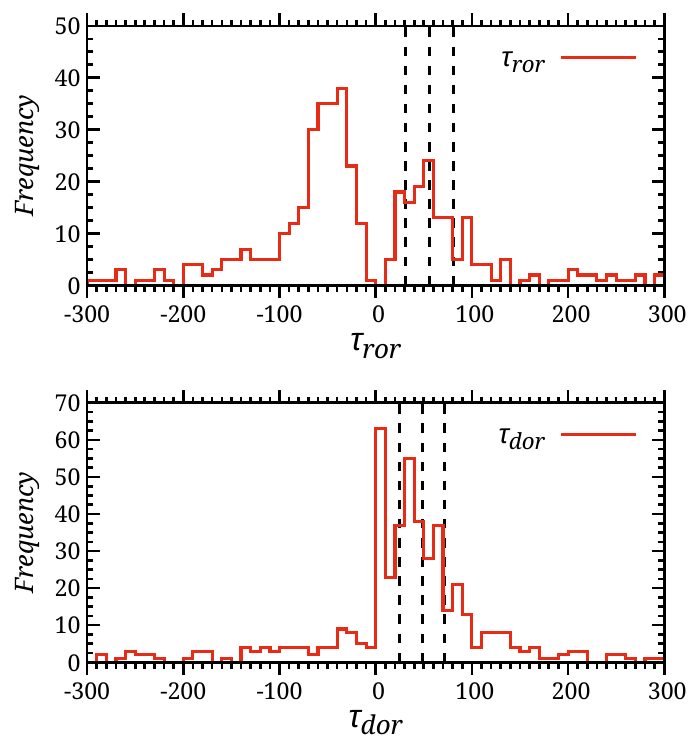}
    \caption{
	Bottom plot: the distribution of all the fitted $\tau_{\scriptscriptstyle dor}$ between -300 to 300 in the fiducial kernel size case. The three vertical dashed lines from left to right show the $\mu$-$\sigma$, $\mu$,  $\mu$+$\sigma$ values in the fitted Student's t-distribution using only the positive values of $\tau_{\scriptscriptstyle dor}$.
	Top plot: the distribution of all the fitted $\tau_{\scriptscriptstyle ror}$ between -300 to 300 in the fiducial kernel size case. The three vertical dashed lines show the $\mu$-$\sigma$, $\mu$,  $\mu$+$\sigma$ values in the fitted Student's t-distribution using only the positive $\tau_{\scriptscriptstyle ror}$.
    }
    \label{fig:11}
\end{figure}

One of the main assumptions of this work is that the distribution of exoplanets is uniform in most subspaces in the $dor$-$ror$ space.
However, the true distribution of exoplanets in the $dor$-$ror$ space is unclear.
There are even theoretical studies and observations showing that most terrestrial exoplanets in close-in orbits have a radius of 1 to 2 $R_{\oplus}$.
Therefore, the assumption of uniform distribution in this paper challenges the reliability of the results.

In the hypothetical case of a 100\% detection efficiency, i.e., there is no observational bias, the distribution density profiles of terrestrial planets along the $dor$-axis and $ror$-axis will show their intrinsic characteristics. 
Assuming that these density profiles in different subspaces can be various situations, including increase, decrease, or constant.
Then if described by exponential decay functions of Equation \ref{eq:taudor} and \ref{eq:tauror}, the distribution of the decay constant of these intrinsic density profiles, named as $\tau_{\scriptscriptstyle dor\_intr}$ and $\tau_{\scriptscriptstyle ror\_intr}$, will range from negative to positive values.

In the case with a realistic detection efficiency profile, the decrease of the detection efficiency at a larger orbital distance and at a smaller planetary radius will result in positive values of decay constant if described by Equation \ref{eq:taudor} and \ref{eq:tauror}.
Accordingly, the distribution of $\tau_{\scriptscriptstyle dor\_intr}$ and $\tau_{\scriptscriptstyle ror\_intr}$ will have an overall positive displacement and become $\tau_{\scriptscriptstyle dor}$ and $\tau_{\scriptscriptstyle ror}$.

Consider two scenarios.
One is that the intrinsic distribution of terrestrial planets in $dor$-$ror$ space is uniform in most subspaces, and the corresponding distribution of $\tau_{\scriptscriptstyle dor\_intr}$ or $\tau_{\scriptscriptstyle ror\_intr}$ approximate a normal distribution with a mean value of 0.
The other is that the intrinsic distribution of planets has obvious characteristics, increasing or decreasing along the positive $dor$-axis or the negative $ror$-axis in most subspaces, and thus cannot be simply represented by a normal distribution.
In the first scenario, after adding the positive displacement caused by a realistic detection efficiency profile, the distribution of decay constants change from $\tau_{\scriptscriptstyle dor\_intr}$ or $\tau_{\scriptscriptstyle ror\_intr}$ with a mean of 0 to $\tau_{\scriptscriptstyle dor}$ or $\tau_{\scriptscriptstyle ror}$ with a positive mean.
The magnitude of the displacement depends on the profile of the detection efficiency.
In the second scenario, a realistic detection efficiency profile also leads to a positive displacement of the overall distribution of $\tau_{\scriptscriptstyle dor\_intr}$ or $\tau_{\scriptscriptstyle ror\_intr}$, 
but it will be difficult to infer the displacement based on the newly formed distribution of $\tau_{\scriptscriptstyle dor}$ or $\tau_{\scriptscriptstyle ror}$ because the distribution of $\tau_{\scriptscriptstyle dor\_intr}$ or $\tau_{\scriptscriptstyle ror\_intr}$ is not known in advance, especially in the case that these distributions are multimodal.

Figure \ref{fig:11} shows the distributions of all the fitted $\tau_{\scriptscriptstyle dor}$ and $\tau_{\scriptscriptstyle ror}$ between -300 to 300 in the fiducial kernel size case, and the locations of the $\mu$-$\sigma$, $\mu$,  $\mu$+$\sigma$ values in the fitted Student's t-distribution using only the positive $\tau_{\scriptscriptstyle dor}$ and $\tau_{\scriptscriptstyle ror}$.
For the distribution of $\tau_{\scriptscriptstyle dor}$, it resembles  the first scenario, i.e., a distribution with a positive mean after a displacement caused by a realistic detection efficiency profile.
Note that it is also possible that the overall distribution of terrestrial planets along the $dor$-axis is not very uniform, and thus the original mean of $\tau_{\scriptscriptstyle dor\_intr}$ is not exactly at 0, and then in that case the $\tau_{\scriptscriptstyle dor}$ obtained in this work will deviate from the decay constant that corresponds to the real detection efficiency profile.
While this possibility is relatively small.
Since the focus of this work is on the terrestrial planets within 1 AU and with sizes between 0.5-4 $R_{\oplus}$.
The formation and evolution of such terrestrial planets are very complicated.
From the initial characteristics of protoplanetary disk to the final assembly stage of orbital crossing, too many physical and random processes are involved \citep{Chambers2001,Ida2004,Mordasini2012,Jin2016}.
Under the influence of various stochastic conditions and processes, 
it is more natural for the distribution of terrestrial planets along the $dor$-axis to be generally uniform from the perspective of probability.
Another point to note is that if the distribution of $\tau_{\scriptscriptstyle dor\_intr}$ is a normal distribution with a mean at 0, the dropping of the negative values in the fitting process will lead to a larger mean of $\tau_{\scriptscriptstyle dor}$ compared to the decay constant that corresponds to the real detection efficiency profile.
However, this discrepancy will be small since the Student's t-distribution is used for fitting.

For the fitted distribution of $\tau_{\scriptscriptstyle ror}$, there are two peaks as shown in Figure \ref{fig:11}.
One peak is composed of negative values, and it corresponds to the subspaces of the third category in Figure \ref{fig:8} where the planetary distribution increases along the negative $ror$-axis, although it is more difficult for planet detection in this direction due to the decrease of detection efficiency.
The other peak is composed of positive values, and it corresponds to the subspaces of the first category in Figure \ref{fig:7} where the planetary distribution decreases along the negative $ror$-axis.
Thus, the $\tau_{\scriptscriptstyle ror}$ distribution resembles the second scenario, i.e., adding a realistic detection efficiency profile results in a positive displacement of the overall distribution of $\tau_{\scriptscriptstyle ror\_intr}$.
Combining Figure \ref{fig:7} and Figure \ref{fig:8}, the overall trend of planetary distribution along the negative $ror$-axis is first increase and then decrease. 
The magnitude of the increase is definitely stronger than that shown in Figure \ref{fig:8}, because the negative impact of the decay of detection efficiency in this direction must be considered. 
Whether the magnitude of decrease is purely due to the decay of detection efficiency or a combined effect of the detection efficiency and an intrinsic declining distribution of planets is not known, since the $\tau_{\scriptscriptstyle ror\_intr}$ distribution on the positive axis in Figure \ref{fig:11} is not clear in advance. 
If the intrinsic planet distribution decreases or slightly increases 
along the negative $ror$-axis, the $\tau_{\scriptscriptstyle ror}$ obtained in this work will be larger or smaller than the decay constant that corresponds to the real detection efficiency profile.
This difference should be small,
because the interval in the $ror$-axis is tiny in the fitting of $\tau_{\scriptscriptstyle ror}$, and correspondingly there should be little change in planetary distribution in such a small radius interval.
It is also worth noting that the dropping of the negative values in the fitting process will lead to a larger mean of $\tau_{\scriptscriptstyle ror}$ compared to the decay constant that corresponds to the real detection efficiency profile, but it has to be done due to the significant distribution peaks of $\tau_{\scriptscriptstyle ror}$ at negative values as shown in Figure \ref{fig:11}. 

The main purpose of this paper is to give the overall distribution characteristics of terrestrial planets of 0.5-4 $R_{\oplus}$ within $\sim$ 1 AU.
The $\tau_{\scriptscriptstyle dor}$ and $\tau_{\scriptscriptstyle ror}$ derived in this work may not be accurate, so Figure \ref{fig:10} also shows the relative occurrence rates generated using the $\tau_{\scriptscriptstyle dor}$ and $\tau_{\scriptscriptstyle ror}$ at the one standard derivations above and below the mean.
These results show that the overall distribution characteristics that correspond to extreme values of $\tau_{\scriptscriptstyle dor}$ and $\tau_{\scriptscriptstyle ror}$ do not changed significantly.
For example, planets with sizes of 0.5-2 $R_{\oplus}$ show the largest relative occurrence rates, and there are relatively few planets with sizes of 2-4 $R_{\oplus}$ within 0.2 AU, etc.

\section{Conclusions}
\label{sec:concl}

In this work, the relative occurrence rate distribution of terrestrial planet is derived from the perspective of data analysis.
I select from the Kepler DR25 data release the information that are related to the following properties of planet candidates: the orbital inclination, the planet-star distance over star radius ($dor$),  and the planet-star radius ratio ($ror$). 
I first use the planetary orbital inclination to correct the geometric factor that affect observability, and then fit two exponential decay functions of detection efficiency along the positive $dor$-axis and the negative $ror$-axis, and finally compensate the distribution density of planet candidates in the $dor$-$ror$ space for the exponential decay functions of detection efficiency to obtain the relative planet occurrence rates. % of terrestrial planet.

The obtained relative occurrence rate distribution of terrestrial planets shows the following features:
\begin{enumerate}
	\item[1.] There is a vacancy of planets with radii between 2.0 and 4.0 $R_{\oplus}$ inside of $\sim$ 0.5 AU, and  the lower boundary of this vacancy gradually increases as the orbital semi-major axis increases. The probable reason for the vacancy may be due to the effect of atmospheric evaporation as proposed by theoretical studies, since the ability to strip planetary envelope decreases as the orbital semi-major axis increases.
	\item[2.] There are two regions with planet abundance. One corresponds to planets with radii between 0.5 and 1.5 $R_{\oplus}$ inside of 0.2 AU, the other corresponds to planets with radii between 1.5 and 3 $R_{\oplus}$ beyond 0.5 AU. The relative occurrence rates in the outer high occurrence region beyond 0.5 AU is about 10 times higher than the inner one inside of 0.2 AU. This indicates that terrestrial planets in the habitable zone around FGK stars are even more common than in close-in orbits.
	\item[3.] The region with planetary radii between 1.5 and 3.0 $R_{\oplus}$ and orbital semi-major axes between 0.2 and 0.5 AU show occurrence rates that are only about one-tenth of the region beyond 0.5 AU, where the occurrence rates are supposed to be lower since the detection efficiency is much lower there. This means that the region between 0.2 and 0.5 AU is a valley with relatively lower planet occurrence rates. This valley corresponds to the evaporation valley that is proposed by theoretical research. Our result shows this valley is not an area of planet vacancy, as there are still considerable planet occurrence rates in this region.
	\item[4.] The relative planet occurrence rates obtained by compensated for extremely fast or slow exponential decay functions of detection efficiency show that the high and low occurrence regions of terrestrial planets do not change much. The main differences in extreme cases are that the relative magnitudes of the two high occurrence rate regions and the low occurrence rate valley change a lot.
\end{enumerate}

This work gives the relative distribution of planets in the planetary radius versus orbital semi-major axis space.
It shows the abundance of terrestrial planets at close-in orbits and at the habitable zone. 
It confirms the distribution features that may be caused by the effect of planetary atmospheric escape as proposed by theoretical studies.
Since the relative planet occurrence rates are normalized in a way that the maximum value is 1, it cannot gives the absolute occurrence rates of different types of planets per star.
The main purpose of this work is to provide a useful constraint for the overall distribution of terrestrial planets inside of 1 AU around FGK stars.
In the future, accurate radii yield from TESS \citep{Ricker2015}, CHEOPS \citep{Broeg2013}, and PLATO 2.0 \citep{Rauer2014} will provide better understanding of the characteristics of terrestrial planets distribution in the planetary radius versus orbital semi-major axis space.

\section*{Acknowledgements}

I thank the anonymous referee for the constructive comments that greatly improved this study.
This work is supported by the B-type Strategic Priority Program of the Chinese Academy of Sciences (Grant No. XDB41000000), National Natural Science Foundation of China (Grant Nos. 11973094, 12033010, 11633009), Youth Innovation Promotion Association CAS (2020319), and Foundation of Minor Planets of Purple Mountain Observatory.
This research has made use of the NASA Exoplanet Archive, which is operated by the California Institute of Technology, under contract with the National Aeronautics and Space Administration under the Exoplanet Exploration Program.

%%%%%%%%%%%%%%%%%%%%%%%%%%%%%%%%%%%%%%%%%%%%%%%%%%

\section*{DATA AVAILABILITY}
The data underlying this article is available upon request to the corresponding author.



%\bibliographystyle{mnras}


%%%%%%%%%%%%%%%%%%%%%%%%%%%%%%%%%%%%%%%%%%%%%%%%%%

% Don't change these lines
\bsp	% typesetting comment
\label{lastpage}
\end{document}